\documentclass[aps,pre,floats,floatfix,twocolumn]{revtex4}

\usepackage{ifthen}
\usepackage{ifpdf}
\usepackage{color}

\ifpdf
\usepackage{graphicx}
\usepackage{epstopdf}
\else
\usepackage{graphicx}
\usepackage{epsfig}
\fi

\usepackage{latexsym}
\usepackage{amsmath}
\usepackage{amssymb}
\usepackage{bm}
\usepackage{wasysym}

\newcommand{\eexp}{\mbox{e}^}

\newcommand{\tbox}[1]{\mbox{\tiny #1}}
 
\newcommand{\mylabel}[1]{\label{#1}} 
\newcommand{\beq}{\begin{eqnarray}}
\newcommand{\eeq}{\end{eqnarray}} 
\newcommand{\be}[1]{\begin{eqnarray}\ifthenelse{#1=-1}
{\nonumber}{\ifthenelse{#1=0}{}{\mylabel{e#1}}}}
\newcommand{\ee}{\end{eqnarray}} 

\newcommand{\Eq}[1]{\textcolor{blue}{Eq.\!\!~(\ref{#1})}} 
\newcommand{\Fig}[1]{\textcolor{blue}{Fig.}\!\!~\ref{#1}} 
\newcommand{\hide}[1]{}
\newcommand{\rmrk}[1]{#1}
\renewcommand{\cite}[1]{\textcolor{blue}{[\onlinecite{#1}}]} 

\begin{document}

\title{Diffusion in sparse networks: linear to semi-linear crossover}

\author{Yaron de Leeuw and Doron Cohen}

\affiliation{\mbox{Department of Physics, Ben Gurion University of the Negev, Beer Sheva 84105, Israel}} 

\begin{abstract}
We consider random networks whose dynamics is
described by a rate equation, with transition rates $w_{nm}$
that form a symmetric matrix. The long time evolution
of the system is characterized by a diffusion coefficient~$D$.
In one dimension it is well known that $D$ can display an abrupt
percolation-like transition from diffusion (${D>0}$)
to sub-diffusion (${D=0}$). A question arises whether
such a transition happens in higher dimensions.
Numerically $D$ can be evaluated using a resistor network
calculation, or optionally it can be deduced from 
the spectral properties of the system. Contrary to a recent 
expectation that is based on a renormalization-group analysis, 
we deduce that $D$ is finite; 
suggest an ``effective-range-hopping" procedure to evaluate it;
and contrast the results with the linear estimate.
The same approach is useful for the analysis of 
networks that are described by quasi-one-dimensional  
sparse banded matrices. 
\end{abstract}

\maketitle


\section{Introduction}

The study of network systems is of interest in 
diverse fields of Mathematics, Physics, and Computer and Life Sciences. 
\rmrk{Commonly a network is described} by a symmetric matrix 
that consists of real non-negative elements, e.g. the adjacency matrix, 
or the link probability matrix, 
that have unique spectral properties \cite{dorogovtsev,bradde}. 
Physically motivated, in this work we consider $d$-dimensional 
network systems, whose dynamics is described by a rate equation
\be{1}
\frac{dp_n}{dt} \ \ =  \ \ \sum_m w_{nm} p_m 
\eeq
The off-diagonal elements of $\bm{w}$ 
are the transition rates, while the diagonal 
elements are the decay rates 
\beq
w_{nn} =-\gamma_n, 
\ \ \ \ \  \gamma_n\equiv \sum_{m (\neq n)} w_{mn}  
\eeq
We assume a symmetric matrix and write schematically
\beq
\bm{w} \ \ = \ \ \text{matrix}\{w_{nm}\}
\eeq
\rmrk{In some sense, one can regard $\bm{w}$ as 
a discrete {\em Laplacian} that is associated with the network.}
Clearly the physical problem is related to the study 
of random walk in a disordered environment \cite{bouch,Montroll,Kaya}.

For presentation purposes we regard the nodes 
of the network as {\em sites}, 
each having a location~$x_n$. 
By construction, we assume that the transition rates $w_{nm}$ 
are given by the expression ${w_0 \eexp{-\epsilon_{nm}} B(x_n{-}x_m)}$,  
where $B(r)$ describes the systematic dependence 
of the coupling on the distance between the sites, 
and $\epsilon$ is a random variable that might  
represent, say, the activation energy that is required 
to make a transition. \rmrk{Consequently the 
network is characterized by two functions:}
\be{4}
w(r,\epsilon) \ \ &\equiv& \ \ w_0 \ \eexp{-\epsilon} \ B(r) \\
\rho(r,\epsilon) \ \ &\equiv& \ \ \mbox{local density of sites}
\eeq
The latter is defined as the density 
of sites in $(r,\epsilon)$ space, 
relative to some initial site. 
Obviously the functional dependence 
of this density on $r$ is affected 
by the dimensionality of the network.

{\em Sparsity.-- }
Our interest is focused on ``sparse" networks.
This means that the transition rates between 
neighboring sites are log-wide distributed    
as in ``glassy" systems. These rates span 
several orders of magnitudes as determined 
by the dispersion of $r$ 
or by the dispersion of $\epsilon$. 
In particular (but not exclusively) we are interested 
in a random site model where the rates depend exponentially 
on the distance between randomly distributed sites, 
namely ${B(r) = \exp(-r/\xi)}$. 
In this particular case one can characterize the 
sparsity by the parameter 
\be{101}
s \ \ = \ \ {\xi} / {r_0}
\eeq
where $r_0$ is the average distance between 
neighboring sites. We refer to such networks 
as ``sparse" if $s\ll1$.

{\em Sparsity vs percolation.-- }
The problem that we consider is a variant of the percolation problem \cite{aa0}:
Instead of considering a bi-modal distribution (``zeros" and ``ones")
we consider a log-wide distribution of rates \cite{Halp}, 
for which the median is much smaller than the mean value. 
We call such a network ``sparse" (with quotation marks) because 
the large elements constitute a minority.

{\em Sparsity vs disorder.-- }
While the standard  ``percolation" problem 
can be regarded as the outcome of extreme ``sparsity", 
the latter can be regarded as arising from an extreme ``disorder".  
Accordingly, the model that we are considering 
is a close relative of the Anderson localization 
problem, and therefore we shall dedicate 
some discussion to clarify the relation.

{\em Physical context.-- }
The model that we address is related and motivated  
by various physical problems, for example: 
phonon propagation in disordered solids \cite{phn1,phn2,amir}; 
Mott hopping conductance \cite{mott,miller,AHL,Halp,pollak,VRHbook};
transport in oil reservoirs \cite{aa1,aa2};
conductance of ballistic rings \cite{kbd};
and energy absorption by trapped atoms \cite{kbw}. 
Optionally these models can be fabricated by combining oscillators: 
say mechanical springs or electrical resistor-capacitor elements. 
In all these examples the issue is to understand how 
the {\em transport} is affected by the {\em sparsity} of a network.  
If the rates are induced by a driving source, 
this issue can be phrased as  going {\em beyond} 
the familiar framework of linear response theory (LRT), 
as explained below.

{\em Diffusion and subdiffusion.-- } 
Our interest is focused on the diffusion coefficient $D$ that characterizes the 
long time dynamics of a spreading distribution. The simplest way 
to define it, \rmrk{as in standard textbooks}, is via 
the variance $S(t) \equiv \left\langle r^2\right\rangle_t$. Namely,
\be{111}
D \ \ \equiv \ \ (2d)^{-1} \ \lim_{t\rightarrow\infty} \frac{S(t)}{t}, 
\eeq 
Optionally it can be defined or deduced from the decay of the 
survival probability ${\mathcal{P}(t) \sim (D t)^{-d/2}}$. 
Hence it is related to the spectral properties of the transition rate matrix. 

In the $d{=}1$ case, it is well known \cite{alexander} that $D$ can display an abrupt 
percolation-like transition from diffusive (${D>0}$) to sub-diffusive (${D=0}$) 
behavior, as the sparsity parameter drops below the critical value ${s_{cr}=1}$.
Similar anomalies are found for fractal structures with ${d<2}$, 
also known as ``random walk on percolating clusters", 
see \cite{havlin1,granek1,granek2,havlin,klemm}. 
A question arises whether such a transition might happen in higher dimensions.

In \cite{amir} the spectral properties in the $d{=}2$ case 
were investigated: on the basis of the renormalization group (RG) procedure 
it was deduced that $\mathcal{P}(t)$ decays in a logarithmic way, 
indicating anomalous (sub) diffusion.  
In the present work we shall introduce a different approach 
that implies, contrary to the simple RG treatment, 
that in spite of the sparsity, the long time dynamics 
is in fact diffusive rather than sub-diffusive.

{\em Resistor network picture.-- }
One can regard the $p_n$ in \Eq{e1} as the charge in site~$n$; 
each site is assumed to have unit capacitance; 
hence $p_n{-}p_m$ is the potential difference; 
and $w_{nm}(p_m{-}p_n)$ is the current 
from~$m$ to~$n$. Accordingly \Eq{e1} can be regarded
as the Kirchhoff equation of the circuit.
While calculating $D$ it is illuminating to exploit 
the implied formal analogy with a resistor network 
calculation \cite{miller,pollak,kbd,camboni}.  
Namely, regarding $w_{nm}$ as connectors, 
it follows that $D$ is formally like conductivity.
It follows that $D[\bm{w}]$ is in general a {\em semi-linear} function: 
\be{180}
D[\lambda \bm{w}] \ &=& \ \lambda D[\bm{w}] 
\\ \label{e181}
D[\bm{w}^a+\bm{w}^b] &>& D[\bm{w}^a]+D[\bm{w}^b]
\eeq
If the rates are induced by a driving source, 
the above super additivity implies that 
the analysis should go {\em beyond} the familiar framework 
of linear-response theory \cite{slk}.

In this work we obtain an improved estimate for~$D$ 
that we call effective range hopping (ERH).
Using this approach we show that in the $d{=}2$ case, as $s$ becomes small, 
the functional $D[\bm{w}]$ exhibits a smooth crossover from ``linear" behavior  
to ``semi-linear" VRH-type dependence.  
Our approach is inspired by the resistor network picture 
of \cite{miller,AHL,Halp,pollak,VRHbook,aa1,aa2,kbd,kbw,slk},  
and leads in the appropriate limit to the well known 
Mott's variable range hopping (VRH) estimate for~$D$.

{\em Outline.-- } We first describe some known 
results, and some additional numerical results, 
for the spectral properties of $d{=}1$ and $d{=}2$ 
networks, and for the dependence of $D$ on the sparsity.
Then we show that an ERH procedure 
is useful in describing the crossover from the 
linear regime (no sparsity) to the semi-linear regime.
In the latter regime a ``resistor network"  approach 
is essential, and the percolation threshold manifests 
itself in the calculation. 
Finally we demonstrate that the same ERH procedure can 
be applied in the case of a quasi-one-dimensional network 
that is described by a sparse banded random matrix. 
The latter is of relevance to previous studies 
of energy absorption by a weakly chaotic system \cite{slk}.
We conclude with a discussion and a short summary.

\section{The random site hopping model}

\rmrk{Consider a network that consists of sites 
that are distributed in space}, locations $x_n$.
With each bond $nm$ we associate an activation 
energy $\epsilon_{nm}>0$, and assume   
\beq
w_{nm} \ \ = \ \  w_0 \ \eexp{-\epsilon_{nm}} \ \eexp{-|x_n-x_m|/\xi} 
\eeq
Accordingly we have the identification 
\beq
B(r) \ \ = \ \ \eexp{-r/\xi} 
\eeq
We note that in the traditional formulation 
of the Mott problem the ``activation energies"
are not due to some ``barriers", 
but are determined by the on-site binding energies, 
namely $\epsilon_{nm}=|\varepsilon_n-\varepsilon_m|/T$,   
where $T$ is the temperature.
In this paper we treat the $\epsilon_{nm}$ 
as an uncorrelated random variable.

The density of sites relative to some initial site 
is characterized  by a joint distribution function 
\be{7}
\rho(r,\epsilon)drd\epsilon = \frac{\Omega_d \, r^{d-1}dr}{r_0^{d}} \ f(\epsilon)d\epsilon,     
\ \ \ \ \ \Omega_d=2,2\pi,4\pi
\eeq
We distinguish between the ``Mott hopping model" 
and the ``degenerate hopping model". Namely, 
\be{72}
f(\epsilon) &=& 1  \ \ \ \ \ \ \ \ \ \ \ \ \mbox{Mott hopping model}   \\        
\label{e71}
f(\epsilon) &=& \delta(\epsilon) \ \ \ \ \ \ \ \ \ \mbox{Degenerate hopping model}
\eeq
The normalization of $f(\epsilon)$ as defined above 
fixes the value of the constant $r_0^d$, which we regard 
as the ``unit cell". 
In the numerics we set the units of distance such that ${r_0=1}$.

In the traditional formulation of the Mott problem 
it is assumed that mean level spacing within $\xi^d$ 
is $\Delta_{\xi}$, such that the number of accessible sites 
is ${(d\varepsilon /\Delta_{\xi}) \, (d^3r/\xi^d)}$. 
\rmrk{By the convention} of \Eq{e7} this implies 
that the unit cell dimension is temperature dependent
\be{100}
r_0^d \ \ = \ \  \left(\frac{\Delta_{\xi}}{T}\right) \, \xi^d
\ \ \ \ \ \ \rmrk{\text{[for Mott model]}}
\eeq
We re-emphasize that the number of sites per unit volume
in the Mott problem is infinite, but effectively 
only $\sim T/\Delta_{\xi}$ sites are accessible 
within $\xi^d$ per attempted transition. 
It is convenient to characterize a random 
site model by a ``sparsity" parameter that 
is defined as in \Eq{e101}. Accordingly 
\be{1011}
s \ \ \equiv \ \ \frac{\xi}{r_0} \ \ = \ \ \left(\frac{T}{\Delta_{\xi}}\right)^{1/d} 
\ \ \ \ \ \ \rmrk{\text{[for Mott model]}}
\eeq
We refer to a network as ``sparse" if $s\ll1$.

\rmrk{The lattice model with near-neighbor (n.n.) transitions} 
is one of the most popular models in statistical mechanics: 
in particular the random walk problem on a lattice 
is a standard textbook example. If the rates are 
generated from a log-wide distribution, it can be regarded 
as a variant of the random site hopping model. 
For details see Appendix \ref{aLattice}. 
In particular we note that the $d{=}1$ version is 
formally equivalent: it does not matter whether 
the distribution of $w$ is due to random distances $r$, 
or due to random activation energies $\epsilon$.  

Finally we note that a quasi-one-dimensional version 
of the random site model arises in the study 
of energy absorption as explained in Appendix \ref{aBanded}, 
and later addressed in Section \ref{sBanded}.

\section{The characterization of transport}

The long time dynamics that takes place on the network 
is characterized by the spreading $S(t)$, 
and by the survival probability $\mathcal{P}(t)$.
If the system is diffusive, these functions
have the following functional form: 
\beq
S(t) &=& \left\langle r^2\right\rangle_t \quad\sim\quad  (2d)Dt\ \\
\mathcal{P}(t) &\sim&  \frac{r_0^d}{\left({4\pi D t}\right)^{d/2}} 
\eeq
See Appendix \ref{diff} for details. The diffusion 
coefficient~$D$ appears here \rmrk{in consistency with 
its definition} in \Eq{e111}. 
\rmrk{We note that in the case of sub-diffusion} 
\beq
S(t) \ \ \propto \ \ t^{\alpha}, 
\ \ \ \ \ \ \ [\alpha<1]
\eeq
which implies by \Eq{e111} that ${D=0}$.

\rmrk{The spectrum} of the matrix $\bm{w}$ consists 
of the trivial eigenvalue $\lambda_0=0$ that is 
associated with a uniform distribution,   
and a set of negative numbers $-\lambda_k$
that describe the decaying modes.   
The spectral function~$\mathcal{N}(\lambda)$ 
counts the number of eigenvalues up to the value~$\lambda$.
We normalize it per site such that ${\mathcal{N}(\infty)=1}$.
The associated density of eigenvalues $g(\lambda)$ 
is related to $\mathcal{P}(t)$ by a Laplace transform.
See Appendix \ref{diff} for details.
It follows that in the case of a diffusive system 
\be{6}
\mathcal{N}(\lambda) \ \ = \ \ \int^\lambda g(\lambda)d\lambda  
\ \ \sim \ \ \left(\frac{r_0}{2\pi}\right)^d\left[\frac{\lambda}{D}\right]^{d/2} 
\eeq
\rmrk{In Appendix \ref{deb} we clarify that this 
expression agrees with {\em Debye law}. 
Accordingly the calculation of $D$ parallels 
the calculation of the speed of sound~$c$ in Debye model.}

Regarded as a transport coefficient $D$ relates 
the probability current to the density gradient.
This is known as Fick's law. 
From the discussion in the Introduction
it follows that $D$~is like the {\em conductivity} 
of a resistor network, which relates the 
electrical current to the voltage difference. 
Some further details on the practical calculation 
of the conductivity are presented in Appendix~\ref{res}. 
On the basis of this analogy it should be clear 
that $D[\bm{w}]$ is in general a {\em semi-linear} 
function of the rates, see \Eq{e181}.

\section{Exact and numerical results for the $d{=}1$ lattice model}
\label{onedim}

In the case of a $d{=}1$ lattice model with n.n. transitions 
it is natural to use the notation $w_n=w_{n,n{-}1}$. 
Pointing out the analogy with adding connectors in series 
the expression for $D$ is 
\beq
D \ = \ \left( \frac{1}{N} \sum_n \frac{1}{w_n} \right)^{-1} 
= \ \ [s>1] \, \frac{s-1}{s} \, w_0
\eeq
The calculation that leads to the last equality 
has been done with the distribution of \Eq{e15}, 
where ${s \equiv \xi/r_0}$.  Note that we have 
here a serial addition of resistors $R=\sum_n R_n$, 
where ${R_n=1/w_n}$.  For ${s<1}$ the distribution 
of each $R_n$ is dominated by the large values, 
hence ${R=\infty}$. On the other extreme for ${s>1}$ 
the distribution of the $R_n$ has finite first and second moments,  
and accordingly the result for $R$ becomes self-averaging, 
as implied by the central limit theorem.
This means the $D$ is ``well defined" only for ${s>2}$. 
For ${1<s<2}$ the result for the average $R$ is finite 
but not self-averaging.

\begin{figure}[h!]
(a) \hspace{0.5\hsize} (b) \\
\includegraphics[height=7cm]{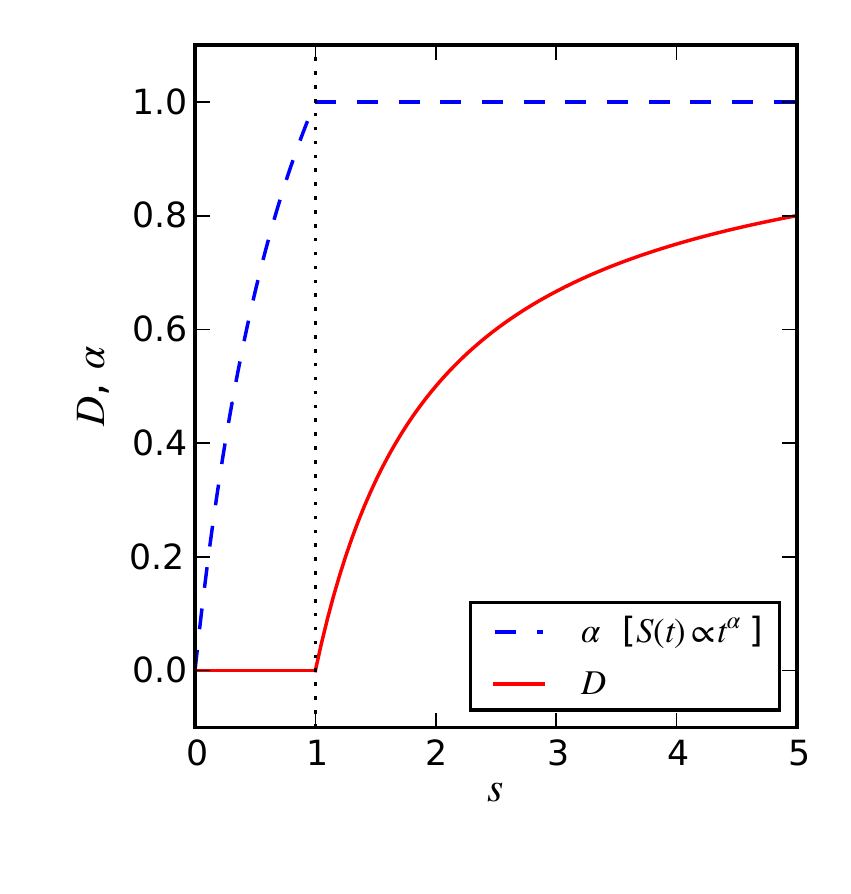}
\includegraphics[height=7cm]{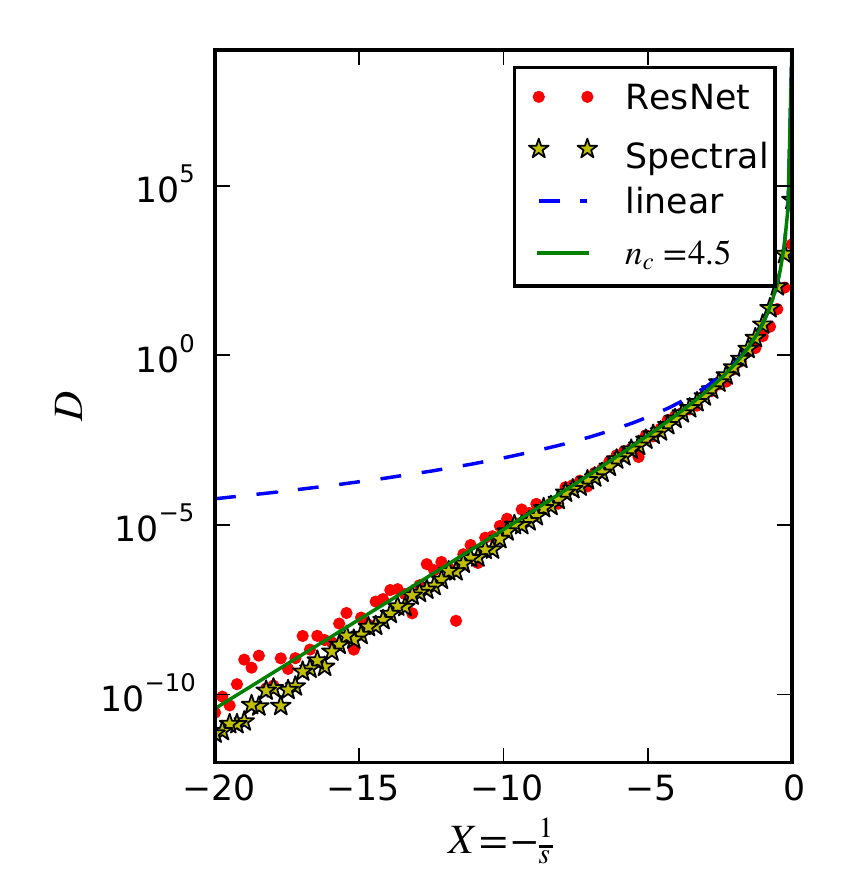}

\caption{Spreading in the $d{=}1$ lattice model (a), 
and in the $d{=}2$ degenerate random site model (b). 
Panel~(a) is based on known exact results.   
Its dashed blue line is the power $\alpha$ of the spreading,  
showing a sub-diffusive regime for $s<1$, and a diffusive regime for $s>1$. 
Its solid red line is the diffusion coefficient $D$, 
which is zero in the sub-diffusive regime. 
Panel~(b) displays numerical results that refer 
to a network that consists of ${N=2000}$ sites 
randomly scattered over a square with periodic boundary conditions.
The vertical axis is the diffusion coefficient $D$ 
in a logarithmic scale, while the horizontal axis is $X=-1/s$. 
The numerical red dots are based on a resistor network calculation (see App.\ref{res}), 
while the stars are extracted from the spectral analysis (see \Fig{f2}).  
The dashed line is the linear estimate (corresponds to ${n_c = 0}$), 
while the solid line is the ERH estimate with ${n_c = 4.5}$.
One observes that the ERH calculation describes very well 
the departure from the linear prediction.
}
\label{f1}
\end{figure}

The dependence of $D$ on $s$ is illustrated in \Fig{f1}a.
In the sub diffusive regime (${s<1}$),  
where the result for the diffusion coefficient is ${D=0}$, 
the dynamics becomes sub-diffusive.  
The explicit results for the survival probability 
and for the spreading are known~\cite{alexander}: 
\beq
S(t) \ \ &\sim& \ \ t^{2s/(1+s)}   \\ 
\mathcal{P}(t) \ \ &\sim& \ \ t^{-s/(1+s)}
\eeq
and the associated spectral function is:
\beq
\mathcal{N}(\lambda) \ \ \sim \ \ \lambda^{s/(1+s)}
\eeq
The numerical demonstration of the latter expectation is displayed 
in \Fig{f2} (left upper panel).
We clearly see that for $s<1$ the asymptotic slope corresponds 
to sub-diffusion, while for ${s>1}$ it corresponds to diffusion.

\section{Numerical results for the $d{=}2$ random site model}

\begin{figure}[h!]
\includegraphics[width=0.7\hsize,clip]{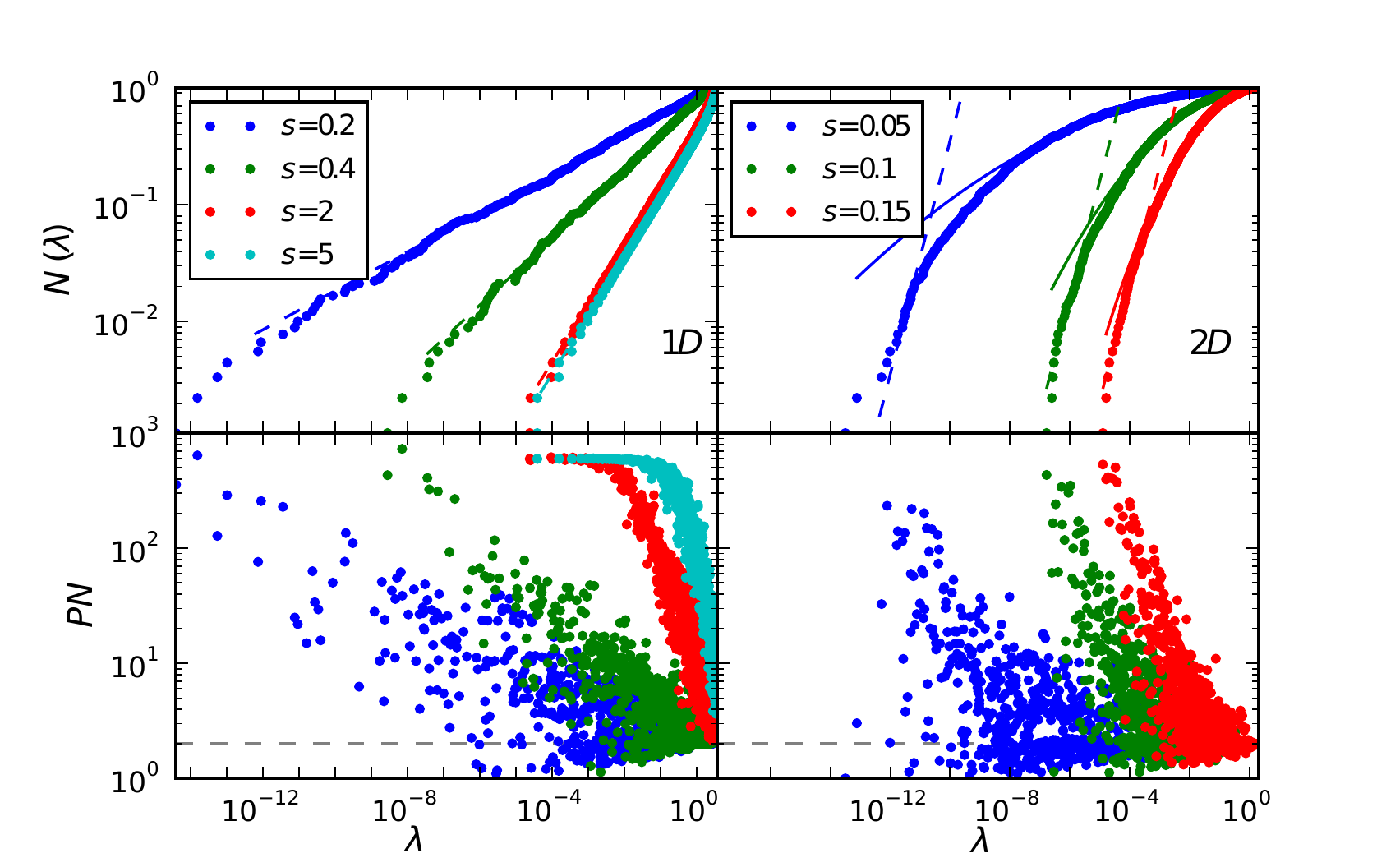}

\caption{ 
The cumulative eigenvalue distributions $\mathcal{N}(\lambda)$ 
for the $d{=}1$ (1D) and for the $d{=}2$ (2D) models of \Fig{f1}, 
and the respective PN of the eigenstates (lower panels).
Several representative values of $s$ are considered.
%
The dots are determined via numerical diagonalization of $N \times N$ matrices,  
each representing a network that consists of ${N=1000}$ sites 
randomly scattered over a square with periodic boundary conditions.
There is a striking difference between the $d{=}1$ and the $d{=}2$ cases. 
For $d{=}1$, the log-log slope of $\mathcal{N}(\lambda)$, 
see dashed lines, is less than~$d/2$ for sparse networks ($s<1$), 
meaning that we have sub-diffusion. 
In the $d{=}2$ case the small-$\lambda$ log-log slope 
is always~$d/2$, which corresponds to normal diffusion.
The solid lines in the upper 2D plot are according 
to the RG analysis of \cite{amir}, namely \Eq{e23}. 
The horizontal dashed line in the lower panels indicates 
the special value PN$=2$ that corresponds to dimer formation.
} 
\label{f2}
\end{figure}

Results for the spectral counting function
of the degenerate $d{=}2$ random site model are presented  
in \Fig{f2} (right upper panel). 
We also display there (in the lower panel) 
the participation number (PN) for each eigenstate. 
The PN of an eigenstate that corresponds to 
an eigenvalue $\lambda_k$ is conventionally 
defined as follows:
\beq 
\text{PN} \ \ \equiv \ \ \left[ \sum_n |\langle n| \lambda_k \rangle|^4 \right]^{-1}
\eeq
As expected from the study of localization
in a disordered elastic medium \cite{loc}, 
the PN becomes larger in the limit ${\lambda\rightarrow 0}$, 
without apparent indication for a mobility threshold.

Assuming localized modes that are conceived via dimerization of 
neighboring sites, $\mathcal{N}(\lambda)$ should equal 
the probability $\exp[-\mathsf{V}(r)/r_0^d]$ 
not to have any neighboring site within 
the volume $\mathsf{V}(r)$ of the sphere ${2w_0 \exp(-r/\xi) > \lambda}$. 
The RG analysis of \cite{amir} refines this naive 
expectation, adding a factor of~2 in the exponent, leading to 
\be{23}
\mathcal{N}(\lambda) \ \ = \ \ \exp\left[ -\frac{\Omega_d}{2d} \Big(-s \ln\left(\frac{\lambda}{2w_0}\right)\Big)^d \right]
\eeq
where $s \equiv \xi/r_0$. 
This expectation is represented in \Fig{f2} (right upper panel) 
by solid lines.  We see that it fails to capture the small $\lambda$ regime, 
where the distribution corresponds to diffusive behavior.

Extracting $D$ via fitting to \Eq{e6} we get \Fig{f1}b. 
We see that in the $d{=}2$ model {\em there is no abrupt crossover 
to sub-diffusion}. We therefore would like to find a way 
to calculate $D$, and hence to have the way to determine 
the small $\lambda$ asymptotics.

\rmrk{{\em Note added.-- }
One should conclude that the RG of reference \cite{amir} 
applies only for the analysis of the high frequency response, 
while our interest is focused in the low frequency (dc) analysis.
The crossover between the two regimes is implied. 
For more details in this direction see a follow-up work \cite{amirNEW}
that confirms our physical picture and demonstrates numerically 
the implied crossover.
}

\section{Linear and ERH estimates for the diffusion coefficient}

The standard way to calculate diffusion
in a $d{=}1$ random walk problem is to inspect 
the transient growth of the variance $\mbox{Var}(n)=2Dt$.
In the stochastic context, if we start at site $n$
we have ${\mbox{Var}(n)=\sum_{n'} p_{n'} (n'-n)^2}$, 
with $p_{n'}=w_{n'n}t$, hence
\beq
D_n \ \ = \ \ \frac{1}{2} \sum_{n'} (n'-n)^2 \ w_{n'n}
\eeq
The generalization to more than one dimension
is straightforward. Averaging the transient expression 
over the starting point we get the result  
\be{25}
D_{\tbox{linear}}  \ \ = \ \ \frac{1}{2d}\iint w(r,\epsilon) \ r^2  \ \rho(r,\epsilon) \ d\epsilon dr 
\eeq
This expression is strictly {\em linear}.
It describes correctly the average transient spreading. 
In the absence of disorder we can trust it for 
arbitrary long time. But if we have a disordered   
or sparse network, the possibility for transport  
is related to the theory of percolation \cite{AHL,Halp,pollak}.
We are therefore motivated to introduce an approximation 
scheme that takes the percolation aspect into account.
We shall refer to this scheme as ``effective range hopping" (ERH) 
because it is a variation on the well known VRH procedure.

Inspired by \cite{AHL,Halp,pollak} we look for the threshold $w_c$ 
that is required for percolation. In the ERH scheme 
we suggest using the following equation for its determination: 
\be{30}
\iint_{w(r,\epsilon)>w_c} \rho(r,\epsilon)drd\epsilon \ \ = \ \ n_c
\eeq
Here $n_c$ is the effective coordination number that is 
required for getting a connected sequences of transitions.
For a $d{=}2$ square lattice model it is reasonable to set $n_c=2$, 
reflecting the idea of forming a simple chain of transitions. 
Rephrased differently the requirement is to have an average of $50\%$ 
connecting bonds per site. 
\rmrk{For a $d{=}2$ random site model one should be familiar 
with the problem of percolation in a system that consists 
of randomly distributed discs.} 
The effective coordination number that is required for getting 
percolation in such a model is ${n_c=4.5}$, as found in \cite{Dalton}, 
and further discussed in Section~IV.A.1 of \cite{Pike}.

The second step in the ERH scheme is to form an effective 
network whose sparse elements are suppressed to the threshold value.  
Then it is possible to use the linear formula \Eq{e25}. Hence we get 

\be{31}
D_{\tbox{ERH}} = \frac{1}{2d}\iint \min\{w(r,\epsilon),w_c\} \ r^2  \ \rho(r,\epsilon) \ d\epsilon dr
\eeq
This expression, as required, is {\em semi-linear} rather than linear. 
It looks like the linear estimate of \Eq{e25}, but it involves a network 
with $w_{nm}$ that are equal or smaller to the original values.
The ``suppressed" connectors are those that are too sparse 
to form percolating trajectories.

\section{Variable range hopping (VRH) estimate}

\begin{figure}[h!]
\includegraphics[width=\hsize]{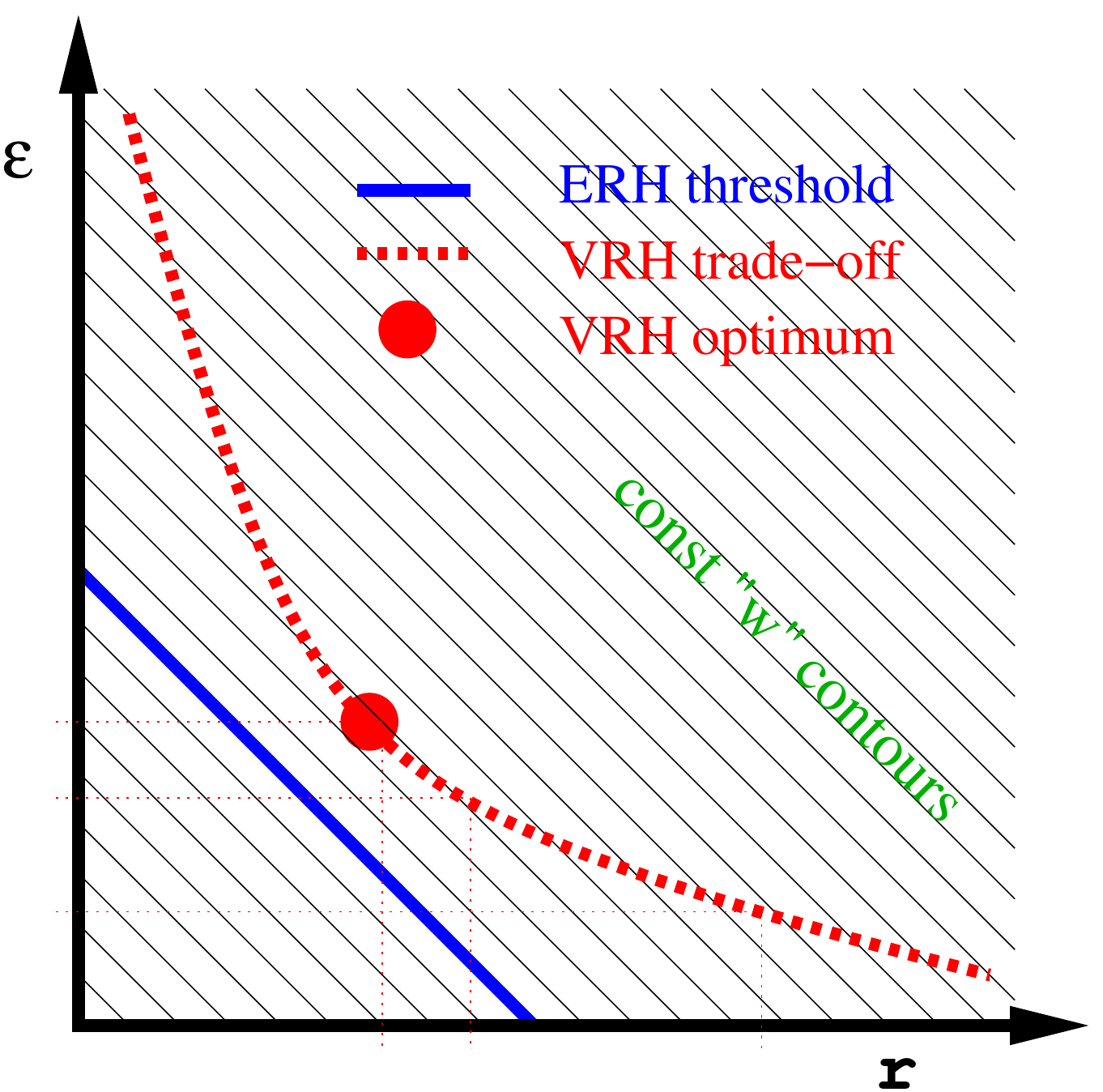}

\caption{Comparing the VRH with the ERH procedure. 
The solid blue line that corresponds to the ERH threshold $w_c$ 
encloses an ``area" that corresponds to $n_c$. 
The VRH trade-off is represented by the dashed red line. 
The VRH optimum is represented by the thick red dot.
The VRH-to-ERH consistency requirement \Eq{e46} 
is to have the VRH optimum sitting on on the solid blue line.}
\label{fv}
\end{figure}

The ERH is similar to the generalized VRH procedure 
that we have used in previous publications \cite{kbd,kbw}.
The traditional VRH is based on the idea of associating
an energy cost $\varepsilon(r)$ to a jump that has range $r$. Namely,  
\be{32}
\varepsilon(r) \ \ \sim \ \ \left[\frac{\Omega_d}{d}\, r^d\right]^{-1} \Delta_0
\eeq
corresponding to the average level spacing 
of the sites within a range~$r$. 
In our notations $\epsilon(r) \equiv \varepsilon(r)/T$.  
For the general network models that we consider here,
the relation between $\epsilon$ and $r$ 
is determined through the equation 
\be{33}
\int_0^{\epsilon} \int_0^{r} \rho(r',\epsilon')  dr'd\epsilon' \  = \  n^* 
\eeq
where $n^*$ is of order unity. In fact we shall deduce later, 
in Section \ref{sM}, that for consistency with the ERH estimate 
this value should be $n^*=n_c/d$. With the substitution 
of \Eq{e7} the trade-off equation can be written as 
\be{34}
\Omega_d\left(\frac{r}{r_0}\right)^d \ F(\epsilon) \ = \ n_c
\eeq
where $F(\epsilon)$ is the cumulative distribution function 
that corresponds to the density $f(\epsilon)$.
In the Mott problem $F(\epsilon)=\epsilon$, and \Eq{e32} is recovered.
In words \Eq{e33} asks what is the $\epsilon$~window that 
is required in order to guarantee that the particle will be able 
to find with probability of order unity an accessible site within 
a range~$r$. Larger jumps allow smaller cost. 
Then we estimate $D$ as follows: 
\be{35}
D_{\tbox{VRH}}  \ \ \sim \ \ 
w^* \times \Big(r^*\Big)^2
\eeq
where $r^*$ is the optimal range that maximizes $w(r,\epsilon(r))$, 
with associated energy cost $\epsilon^*=\epsilon(r^*)$, 
and effective transition rate ${w^*= w(r^*,\epsilon^*)}$. 
See \Fig{fv} for illustration.

The VRH estimate, unlike the ERH, 
does not interpolate with the linear regime. 
It can be used to estimate $D$ 
only if the system is very sparse ($s\ll1$).
It can be regarded as an asymptotic evaluation of 
the ERH integral: it assumes that the hopping is dominated 
by the vicinity of the optimal point ${(r^*,\epsilon^*)}$.
Accordingly, VRH-to-ERH consistency requires the identification ${w^*=w_c}$.  
\rmrk{However, using known results from percolation theory,} 
one possibly can further refine the determination 
of the optimal value $w^*$. Namely, a somewhat smaller 
value than the threshold value $w_c$ might allow a better connectivity.    
As~$s$ becomes very small, the effective range $\delta w$ in the ERH integral, 
which contains the dominant contribution, becomes very small compared with $w_c-w^*$, 
and one should be worried about the implied (sub-dominant) correction. 
This speculative crossover is beyond the scope of the present study, 
and possibly very hard to detect numerically. A useful analogy  
here is with the crossover from ``mean-field" to ``critical" behavior 
in the theory of phase transition, as implied by the Ginzburg criterion.

\section{ERH calculation for the $d{=}2$ lattice model}

The $d{=}2$ lattice model, as defined in Appendix \ref{aLattice}, 
is the simplest and most common example 
for studies of percolation and percolation-related problems. 
We substitute into \Eq{e30} the effective density \Eq{e17}
with the coordination number ${c_L=4}$, and deduce 
that $w_c$ is merely the {\em median} value of the n.n. transition rates. 
%
The ERH calculation using \Eq{e31} with \Eq{e17} 
requires a simple $f(\epsilon)d\epsilon$ integration, 
which can be rewritten as $\tilde{f}(w)dw$ integral.  
This integral is the sum of ${w>w_c}$ and ${w<w_c}$ 
contributions, namely  
\be{37}
D_{\tbox{ERH}} =  \left[\frac{1}{2}w_c + \frac{1}{2}\int_0^{w_c} w\tilde{f}(w) dw\right]r_0^2
\eeq
Note that the first term in the square brackets 
originates from the ${w>w_c}$ contribution.
Note also that the result is $D= w_c r_0^2$ for a delta 
distribution, i.e. in the absence of disorder.

\section{ERH calculation for the degenerate hopping model}
\label{sN}

We now turn to the calculation of the ERH estimate for the
degenerate hopping model. The ERH threshold 
can be written as  $w_c=w_0\exp(-r_c/\xi)$, 
where $r_c$ is determined \rmrk{through} \Eq{e30}, 
which takes the form   
\be{735}
\int_0^{r_c} \frac{\Omega_d r^{d-1} dr}{r_0^2} \ = \  n_c 
\eeq
\rmrk{leading to} 
\beq
w_c &=& w_0\exp\left(-\frac{r_c}{\xi}\right) \\
r_c  &\equiv& \left(\frac{d}{\Omega_c}n_c\right)^{1/d}  r_0
\eeq
\rmrk{The calculation of the ERH integral of \Eq{e31} 
is detailed} in Appendix \ref{aERH}. We note that the 
linear approximation of \Eq{e25} is formally obtained 
by setting $r_c=0$, leading to 
\be{400}
D_{\tbox{linear}} \ \ = \ \  
\frac{(d{+}1)!\,\Omega_d}{2d} \, s^{d{+}2} \, w_0 r_0^2
\eeq
Then it is possible to write the result of the ERH integral as
\be{40}
D_{\tbox{ERH}} \ \ = \ \  
\mathrm{EXP}_{d{+}2}\left(\frac{1}{s_c}\right)  \  \eexp{-1/s_c}  \ D_{\tbox{linear}}
\eeq
where $s_c=\xi/r_c$, and 
\be{41}
\mathrm{EXP}_{\ell}(x) \ \ = \ \ \sum_{k=0}^{\ell} \frac{1}{k!} \, x^k
\eeq
The linear result is formally obtained 
by setting ${n_c=0}$ or in the $d\rightarrow\infty$ limit. 
In the other extreme of ${s\ll1}$ 
we get a VRH-like dependence
\beq
D \ \ \sim \ \ \eexp{-1/s_c}, \ \ \ \ \ \ \ \text{for} \ s\ll1 
\eeq

{\em Numerical verification.-- } 
To obtain an ERH estimate we have to fix  
the parameter $n_c$ in \Eq{e735}. 
One approach is to regard it as a free fitting parameter.
\rmrk{But it is of course better not to use any fitting parameters.}
Fortunately we know from \cite{Dalton,Pike} 
that $n_c=4.5$ is the average number of bonds 
required to get percolation. 
The verification of the ERH estimate for 
the random site model with this value is 
demonstrated in \Fig{f1}(b).

\section{ERH calculation for the Mott hopping model}
\label{sM}

We turn to calculating the ERH estimate for the non-degenerate Mott hopping model, 
and contrast it with the linear approximation, and with the traditional VRH estimate.
The ERH threshold is determined \rmrk{through} \Eq{e30}, leading to   
\beq
w_c &=& w_0\exp(-\epsilon_c) \\ 
\epsilon_c &\equiv& \left( \frac{d}{\Omega_d} \frac{n_c}{s^d} \right)^{1/(d+1)}
\eeq 
In the VRH procedure the optimal hopping range 
is found by maximizing $w(r,\epsilon)$ 
along the trade-off line of \Eq{e34}, 
as illustrated in \Fig{fv}, leading to 
\beq
r^* \ \ = \ \ \left(\frac{d^2}{\Omega_d} n^* \, s \right)^{1/(d+1)} r_0
\eeq 
and the associated rate is 
\be{46}
w^* = w_c \ \ \ \ \ \text{provided} \ n^*=n_c/d
\eeq 
This identification is nexessary if we want the VRH to 
describe correctly the asymptotic dependence of $D$ on $s$.

\rmrk{The calculation of the ERH integral of \Eq{e31} 
is detailed} in Appendix \ref{aERH}. 
Thanks to our conventions the linear result is the same as \Eq{e400},  
and the the final result can be written as follows:  
\be{47}
D_{\tbox{ERH}} \ \ = \ \  
\mathrm{EXP}_{d{+}3}\left(\epsilon_c\right)  \  \eexp{-\epsilon_c}  \ D_{\tbox{linear}}
\eeq 
where $\mathrm{EXP}(x)$ is the polynomial 
defined in \Eq{e41}. 
The linear result is formally obtained by setting ${\epsilon_c=0}$
or in the $d\rightarrow\infty$ limit. 

We see that the VRH estimate can be regarded as an 
asymptotic approximation that holds for ${s\ll1}$. 
Using \Eq{e100} and \Eq{e1011} we deduce 
from  \Eq{e400} and from \Eq{e47} that 
\beq
D_{\tbox{linear}} \ \ \propto \ \ T
\eeq
while for ${s\ll1}$
\beq
D_{\tbox{ERH}} \ \ \sim \ \ \left(\frac{1}{T}\right)^{2/(d+1)} 
\  \exp\left[-\left(\frac{T_0}{T}\right)^{1/(d+1)}\right]
\eeq
where $T_0$ is a constant.

\section{ERH calculation for the banded quasi-one-dimensional model}

\label{sBanded}

We can apply the ERH calculation also to the case 
of the quasi-one-dimensional model that we have 
studied in the past \cite{kbd,kbw}.
This model is motivated by studies of energy absorption \cite{slk}.
For details see Appendix \ref{aBanded}.
The network is defined by a banded matrix $\bm{w}$.
\rmrk{For simplicity} we assume that the sites  are equally 
spaced, and that the reason for the ``sparsity" is the log-wide 
distribution of the in-band elements.

The ERH threshold $w_c$ is deduced from \Eq{e30}. 
For a general $B(r)$ and $f(\epsilon)$ one can integrated 
over $d\epsilon$, and then it takes the form 
\be{50}
\int_0^{\infty}  
\frac{\Omega_d \, r^{d-1}dr}{r_0^{d}} 
\ F\left(\log\left(\frac{w_0}{w_c}B(r)\right)\right) 
\ \ = \ \ n_c
\eeq 
where $F(\epsilon)$ is the cumulative distribution function 
that corresponds to the density $f(\epsilon)$.
Here we are considering a $d{=}1$ network. 
\rmrk{However, we are dealing with a banded matrix  
which in some sense is like adding 
an extra (but bounded) dimension to the lattice.} 
 
Specifically we assume that ${B(r)=1}$ within the band, 
and zero for ${|r|>b}$. 
The non zero elements have a log-box distribution, 
namely, $\epsilon$ is distributed uniformly over a range $[0,\sigma]$. 
To have large $\sigma$ means ``sparsity".
One should notice that this sparsity is less 
traumatic than having $s\ll1$ in the $d{=}1$ lattice model 
that we have considered in Section \ref{onedim}.
This is because the distribution is bounded 
from below by a finite non-zero values. 
Accordingly we cannot have sub-diffusion here.  

\begin{figure}[h!]

(a) 

\includegraphics[width=\hsize,clip]{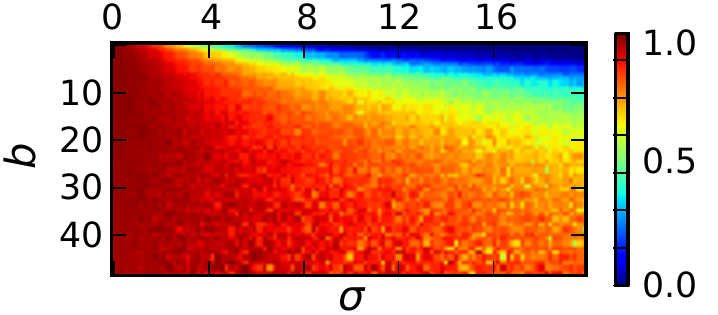}

\ \\ \ \\ \ \\ 

(b)  
\includegraphics[width=\hsize,clip]{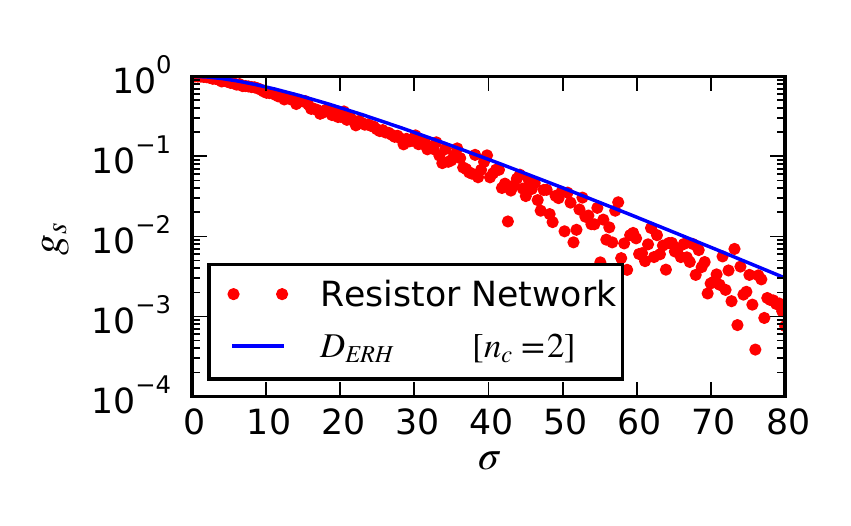}

\ \\ \ \\ \ \\ 

(c) 
\includegraphics[width=\hsize,clip]{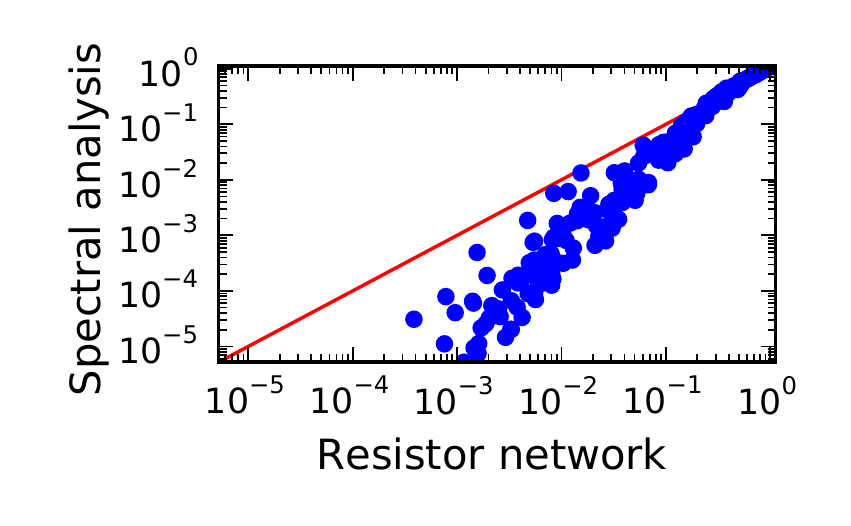}

\caption{
We consider a quasi $d{=}1$ network that consists 
of ${N=1000}$ sites with periodic boundary conditions. 
The network is described by a sparse banded matrix. 
The bandwidth is~$b$, and the log-width of the rate 
distribution is~$\sigma$. See text for details.
{\bf \ (a)}~The numerical result for $g_s=D/D_{\tbox{linear}}$ 
imaged as a function of $\sigma$ and~$b$.
The values of $D$ are found via a numerical 
resistor network calculation, see App.\ref{res}.  
{\bf \ (b)}~Plot of the subset of results that 
refer to the $b=10$ matrix. 
The curve is the ERH prediction. 
{\bf \ (c)}~Scatter diagram that shows 
the correlation between the "$D$" that is extracted 
from the spectral analysis, and the $D$ that  
has been found via the resistor network calculation.}
\label{f4}
\end{figure}

We now turn to estimate $D$ using the 
the ERH procedure. It should be clear that the 
success here is not guaranteed for reasons 
that we further discuss in the last paragraph
of this section.
From \Eq{e50} it follows that ${w_c=w_0\exp(-\epsilon_c)}$, 
where $\epsilon_c$ is the solution of 
\beq
2b \ F\left( \epsilon_c \right) \ \ = \ \ n_c
\eeq
For the assumed $\epsilon$ distribution the 
solution of this equation is trivial 
\beq
\epsilon_c \ \ = \ \ \frac{n_c}{2b}\sigma
\eeq
While doing the ERH integral of \Eq{e31} note that 
the integral $dr$ should be replaced by a sum.
It is convenient to define
\beq
\tilde{b} \ \ \equiv \ \ \sum_{r=1}^b r^2 \ \ = \ \ = \frac{1}{6}b(b+1)(2b+1)
\eeq
Then the ERH estimate takes the form
\beq
D_{\text{ERH}} \ \ = \ \ 
\ \frac{1}{\sigma}\left[ 
\left(1+\frac{n_c}{2b}\sigma\right)\eexp{-\frac{n_c}{2b}\sigma} - \eexp{-2\sigma}
\right] \ \tilde{b} w_0
\eeq 
The linear estimate of \Eq{e25} is formally obtained 
by setting ${n_c=0}$, and in the absence 
of disorder it obviously reduced to $D=\tilde{b} w_0$. 
We define 
\beq
g_s \ \ = \ \ D/D_{\text{linear}}
\eeq
Numerical results are presented in \Fig{f4}, 
and they agree with the ERH estimate.

At this point one wonders whether $D$ can be extracted 
from the spectral analysis, i.e. via fitting to \Eq{e6}.
In \Fig{f4}c we plot the "$D$" that is extracted 
from the spectral analysis versus the $D$ that  
has been found via the resistor network calculation.
We observe that the obtained values are much smaller.
Our interpretation for that is as follows: 
the density of eigenvalues is related to the survival 
probability $\mathcal{P}(t)$ via a Laplace transform;  
For a quasi-one-dimensional system there is a short time $d{=}2$ like 
relatively fast transient; Consequently the $d{=}1$ decay 
holds only asymptotically with a smaller prefactor.  
Accordingly we do not know whether there is a wise 
way to deduce $D$ from the spectral analysis in the 
case of a qausi-one-dimensional network.

Concluding this section we would like to warn the reader
that the use of the percolation picture in $d{=}1$ is 
somewhat problematic: strictly speaking  
there is no percolation transition. Obviously 
for ${b=1}$ we are back with the $d{=}1$ lattice model 
for which there is sub-diffusion if ${s<s_{cr}}$ 
with ${s_{cr}=1}$. However, if $b$ is reasonably large, 
it is not feasible to encounter such an anomaly 
in practice. Even if the distribution is not bounded 
from below, the redundancy due to ${b>1}$ 
would lower the effective value of $s_{cr}$. 
Furthermore: in the Fermi-golden-rule picture (see next section) 
the occurrence of ``weak links" along the band  
are practically not possible because the matrix elements $V_{nm}$ 
are not uncorrelated random variables. 
We can refer to this as the {\em rigidity}.
This rigidity is implied by semi-classical considerations.

\section{Semilinear response perspective}

Considering models of energy absorption, see (Appendix \ref{aBanded}), 
it is assumed that the transition rate $w_{nm}$, between 
unperturbed energy levels $m$ and $n$, is determined 
by a driving source that has spectral content $\tilde{S}(\omega)$.
The Fermi golden rule can be written as  
\beq
w_{nm} \ \ = \ \ \tilde{S}(E_n-E_m) \ |V_{nm}|^2
\eeq
where $V_{nm}$ is the perturbation matrix in the Hamiltonian.
Accordingly we can write instead of~$D=D[\bm{w}]$ 
an implied relation~$D=D[\tilde{S}(\omega)]$.
This relation is in general semi-linear. 
This means that only the first property below, 
which corresponds to \Eq{e180} is satisfied, 
not the second one.
\beq
D\big[ \lambda \tilde{S}(\omega)\big]  &=&   \lambda \ D\big[\tilde{S}(\omega)\big] 
\\
D\big[\tilde{S}_a(\omega) + \tilde{S}_b(\omega)\big]  &=&  D\big[\tilde{S}_a(\omega)\big] + D\big[\tilde{S}_b(\omega)\big]
\eeq
To have a semilinear rather than linear response 
may serve as an experimental signature for the 
applicability of resistor-network modeling of energy 
absorption. We note, however, that if the 
the driving were added ``on top" of a bath, 
the response would become linear at small intensities. 
Namely, if one substituted    
\beq
\tilde{S}(\omega)_{\tbox{total}} \ \ = \ \ \tilde{S}_{\tbox{bath}}(\omega) + \tilde{S}(\omega)
\eeq  
it would be possible to linearize~$D$ with respect 
to the $\tilde{S}(\omega)$ of the driving source. 

The statement that VRH is a ``semilinear response" theory rather than ``linear response"
theory is a source for non-constructive debates on terminology. The reason for the 
confusion about this point is related to the physical context. 
Do we calculate ``current vs bias" or do we calculate ``diffusion vs driving".
The response is linear in the former sense, but semi-linear in the latter sense.

\section{Discussion}

It should be clear that there are two major routes in developing  
a theory for~$D$. Instead of deducing it from spectral properties 
as in \cite{amir}, one can try to find ways to evaluate it directly 
via a resistor network calculation \cite{miller,AHL,Halp,pollak,VRHbook}, 
leading in the standard Mott problem to the VRH estimate for~$D$.   

In \cite{kbd,kbw,slk} this approach was extended 
to handle ``sparse" banded matrices whose elements have 
log-wide distribution, leading to a generalized VRH estimate. 
In this work we have pursued the same direction and obtained an 
improved estimate for~$D$, the ERH estimate.
Using this approach we showed that in the $d{=}2$ case, 
as $s$ becomes small, the functional $D[\bm{w}]$ exhibits 
a smooth crossover from ``linear" behavior to ``semi-linear" VRH-type dependence.  

\rmrk{{\em Relation to other models.-- }}
Disregarding the ``sparsity" issue, the model that we were considering 
is a close relative of the Anderson localization problem.
However it is not the same problem, and there are important 
differences that we would like to highlight.
For the purpose of this discussion it is useful to be reminded 
that the hopping problem that we have addressed is essentially 
the same as studying the spectrum of vibrations in 
a disordered elastic medium. Hence $D^{1/2}$ parallels the  
speed of sound $c$ of the Debye model. See Appendix \ref{deb}.

\rmrk{{\em Mott vs Anderson.-- }}
In the hopping model all the off diagonal elements are positive numbers, 
while the negative diagonal elements compensate them.
It follows that we cannot have ``destructive interference", 
and therefore we do not have genuine Anderson localization. 
Consequently in general we might have diffusion, even in $d{=}1$. 
In $d{=}2$ we have a percolation threshold, which is again 
not like Anderson localization. 
See the discussion of fractons in \cite{havlin}.

\rmrk{{\em Debye vs Anderson.-- }}
In the standard Anderson model the eigenvalues form 
a band ${\lambda \in [-\lambda_c,\lambda_c]}$.
The states at the edge of the band are always localized.
The states in the middle of the band might be de-localized if ${d>2}$.  
The spectrum that characterizes the hopping model does
not have the same properties. With regard to the 
localization of vibrations in a disordered elastic medium \cite{loc},  
it has been found that the spectrum is ${\lambda \in [0,\lambda_c]}$.
The ground state is always the ${\lambda=0}$ uniform state.
The localization length diverges in the limit ${\lambda \rightarrow 0}$.
Consequently the Debye density of states is not violated:
the spectrum is asymptotically the same as that of a 
diffusive (non-disordered) lattice. It follows that the 
survival probability should be like that of a diffusive 
system, and therefore we also expect, and get, diffusive behavior
for the transport: spreading that obeys a diffusion equation.

\section{Summary}

This was originally motivated by the necessity 
to improve the resistor-network analysis of the diffusion 
in quasi-one-dimensional networks \cite{kbd}, and additionally from the 
desire to relate it to the recent RG studies \cite{amir} 
of the spectral properties of random site networks.
The key issue that we wanted to address was 
the crossover from linear-like to semi-linear dependence 
of $D$ on the rates. This crossover show up as the ``sparsity" 
of the system is varied. 

It should be clear that unlike the RG based expectation 
of \cite{amir}, our analysis indicates that there is 
no sub-diffusive behavior in $d{=}2$.
Accordingly, the anomalous $log(t)$ spreading that is 
predicted in \cite{amir} should be regarded as a transient: 
for very small value of the sparsity parameter this transient 
might have a very long duration, but eventually normal diffusion  
takes over. 

One can regard ``sparsity" as an extreme type of disorder:
the rates are distributed over many orders of magnitudes.
Still, unlike the $d{=}1$ case, the implication of ``sparsity" 
in $d{=}2$ is not as dramatic: there is no ``phase transition" 
between two different results, but a smooth crossover.
It is therefore clear that our statements are consistent 
with those of older works that relate to the diverging localization 
properties of the low frequency vibrations in disordered elastic medium \cite{loc}.

The effective range hopping (ERH) procedure that we tested 
in this paper is a refinement of well known studies of variable 
range hopping \cite{mott,miller,AHL,Halp,pollak,VRHbook}.
We used the insight of \cite{AHL,Halp,pollak} that connects VRH 
with the theory of percolation.

Disregarding possible inaccuracy in the determination 
of the optimal rate, the ERH calculation provides 
a {\em lower} bound for $D$. Accordingly, by obtaining 
a non-zero result it is rigorously implied that $D$ is finite. 
The purpose of the numerics was to demonstrate that
in practice the outcome of the ERH calculation provides  
a very good estimate of the actual result, 
interpolating very well the departure from linearity.

It was important for us to clarify that a large class of networks 
can be treated {\em on an equal footing}. In particular we demonstrated that  
the application of the ERH estimate does not require any fitting parameters.
We have verified that the same prescription can be applied both in the case 
of the $d{=}2$ lattice model, and in the case of the $d{=}2$ random-site model, 
provided one uses the appropriate percolation threshold 
that is known from percolation theory.  

For the traditional Mott hopping model 
and its degenerated version we obtained 
the refined expressions \Eq{e47} and \Eq{e40} respectively. 
\rmrk{In these expressions the {\em full} dependence on 
the dimensionality~($d$) is explicit, and the crossover 
to linear response as a function of the sparsity~($s$)
is transparent. Note that in the degenerate random site model 
the sparsity is merely a geometrical feature, while in the 
non-degenerate Mott model the sparsity depends on the temperature 
as implied by \Eq{e1011}.}     

We would like to re-emphasize that the original motivation 
for this work is was the study of energy absorption by driven mesoscopic systems. 
In this context the implication of the semi-linear crossover 
is the breakdown of linear response theory. The latter 
issue has been extensively discussed in past publications \cite{slk}. 

\ \\

{\bf Acknowledgments.-- }
We thank Amnon Aharony, Ariel Amir, Ora Entin-Wohlman, Rony Granek, and Joe Imry 
for illuminating discussions, comments, and references. 
This work has been supported by the Israel Science Foundation (ISF).

\ \\ \ \\ 

\appendix

\section{Lattice model with n.n. hopping}

\label{aLattice}

For $s \ll 1$ the $d{=}1$ random site model is 
essentially equivalent to a lattice model 
with equally spaced sites, 
near neighbor transitions, 
and random $\epsilon$.
From the identification $\epsilon=r/\xi$ 
it follows that the distribution 
of the ``activation energy" is 
\beq
f(\epsilon) \ \ = \ \ s \ \exp(-s\epsilon), 
\hspace{15mm} s \equiv \xi/r_0
\eeq
This implies that the the distribution of the rates is 
\be{15}
\tilde{f}(w) dw  \ \  =  \ \    \ [w < w_0] \, \frac{s \, w^{s-1} dw}{w_0^s}, 
\eeq
The density of sites to which a transition can occur is
%
\be{17}
\rho(r,\epsilon) =  c_L \delta(r-r_0) \ f(\epsilon)  
\eeq 
where $c_L=2$ is the coordination number. 
This corresponds to the $d{=}1$ case of \Eq{e7}.

The $d{=}2$ version of the lattice model has no 
strict relation to the $d{=}2$ random site model. 
A popular choice is to assume a box distribution
for the activation energy within 
some interval ${0<\epsilon<\sigma}$. 
The density of sites to which a transition can occur 
is $2\pi r f(\epsilon)$ for large $r$, 
as implied by \Eq{e7}. But for small $r$  
the effective density is given by \Eq{e17}
with the coordination number ${c_L=4}$.

\section{The quasi-one-dimensional banded matrix model}

\label{aBanded}

On equal footing we consider the quasi-one-dimensional banded lattice model.
This model is motivated by studies of energy absorption \cite{slk}.
In this context the transition rates are determined 
by the Fermi-Golden-Rule (FGR). Hence we write:
\be{143}
w_{nm} \ \ = \ \ w_0 \ \eexp{-\epsilon_{nm}} \ B\left(E_n-E_m\right)
\eeq
Here $n$ and $m$ are unperturbed energy levels of the system, 
but we shall keep calling them ``sites" in order to avoid 
duplicated terminology.
The density of sites relative to some initial site 
is characterized by the same joint distribution function
as for the $d{=}1$ network,  
\beq
\rho(r,\epsilon) \ \ = \ \ 2f(\epsilon) 
\eeq 
Here $r=|E_n-E_m|$ is the distance between the energy levels, 
which is formally analogous to $r=|x_n-x_m|$ in the random site hopping model.
We use here units such that the mean level spacing is unity.
In the later numerical analysis we assume equally spaced levels 
such that the distance is simply ${r=|n-m|}$. 

In the physical context the band profile $B(r)$ is determined 
by the semiclassical limit, while the distribution of the $\epsilon$ 
values is implied by the intensity statistics of the matrix elements.
This intensity statistics is known as Porter-Thomas in the strongly 
chaotic case, corresponding to the Gaussian ensembles, 
but it becomes log-wide for systems with ``weak quantum chaos" \cite{SparseMat},
reflecting the sparsity that shows up in the limiting
case of integrable system \cite{kbw}. 

In the numerical analysis we have considered simple 
banded matrices, for which $B(r)=1$ for ${r \leq b}$, 
and zero otherwise. Accordingly $1{+}2b$ is the bandwidth.
The elements within the band are log-box distributed:
this means that $\epsilon$ is distributed uniformly over a range $[0,\sigma]$.  
Note that log-box distribution is typical of glassy systems, 
where the tunneling rate depends exponentially on the distance  
between the sites.

\section{Numerical extraction of $D$}
\label{diff}

In a diffusive system the coarse grained spreading 
is described by the standard diffusion equation,
with an evolving Gaussian distribution  
\beq
\rho(x;t) \ \ = \ \ \prod_{i=1}^d \frac{1}{\sqrt{2\pi S_x(t)}} 
\exp\left[-\frac{x_i^2}{2S_x(t)} \right]
\eeq
where $S_x(t)=2Dt$. It follows from this expression that 
\beq
S(t) \ \ = \ \ \left\langle r^2(t) \right\rangle  \ \ = \ \ (2d)Dt
\eeq
Starting with all the probability concentrated 
in one ``unit cell" we get for the survival probability 
\beq
\mathcal{P}(t) \ \ \sim \ \ \frac{r_0^d}{\left({4\pi D t}\right)^{d/2}} 
\eeq
The eigenvalues of the diffusion equation are 
\be{323}
\lambda_k \ = \ Dq_k^2, 
\ \ \ \ \ \  k=\text{index}
\eeq
where the possible values of the momentum are determined  
by the periodic boundary conditions as $q=(2\pi/L)\vec{k}$. 
It follows that the cumulative number of eigenstates 
per site is  
\be{324}
\mathcal{N}(\lambda) \ \ = \ \ 
\left(\frac{r_0}{2\pi}\right)^d
\frac{\Omega_d}{d}
\left[\frac{\lambda}{D}\right]^{d/2}
\eeq

It is well known that the survival probability 
is related to the eigenvalues of $\bm{w}$ through the relation
\beq 
\mathcal{P}(t) \ \ = \ \ \frac{1}{N}\sum_\lambda \eexp{-\lambda t} 
\ \ \equiv \ \ \int_0^{\infty} g(\lambda)d\lambda \ \eexp{-\lambda t}
\eeq
For a diffusive system one can verify that 
the expressions above for $g(\lambda)$ and $\mathcal{P}(t)$ 
are indeed related by a Laplace transform.  
More generally, it follows that $D$ can be deduced from 
the asymptotic behavior of $g(\lambda)$
in the ${\lambda\rightarrow 0}$ limit  
where the diffusive description is valid.
In contrast to that for large $\lambda$ we expect $g(\lambda)$ to coincide 
with the distribution of the decay rates $\gamma_n=\sum_{m}w_{mn}$, 
reflecting localized modes.

\section{Relation to Debye model}
\label{deb}

Consider a system of units masses that are connected by springs.
Once can describe the system by a matrix $\bm{w}$ whose 
of-diagonal elements $w_{nm}$ are the spring constants.
The eigen-frequencies are determined accordingly, namely, ${\omega_k = \sqrt{\lambda_k}}$.
Assuming that the low lying modes are like 
acoustic phonons with dispersion ${\omega = c |q|}$,
where $c$ is the so called speed of sound,  
one deduces that 
\beq
\omega_k \ = \ \ = \ c |q_k|, 
\ \ \ \ \ \  k=\text{index}
\eeq 
Consequently the associated counting function is as in the Debye model:
\beq 
\mathcal{N}(\omega) \ \ = \ \ 
\left(\frac{r_0}{2\pi c}\right)^d
\frac{\Omega_d}{d}
\ \omega^{d}
\eeq
Comparing the above expressions with \Eq{e323} and \Eq{e324} 
it follows that the calculation of~$c^2$ 
is formally the same as the calculation of~$D$.

\section{The resistor network calculation}
\label{res}

The diffusion coefficient $D$ is formally like the calculation 
of the conductivity of the network. Therefore it can be determined 
via a numerical solution of a circuit equation.   
It is convenient to use the language of electrical engineering 
to explain how the resistor network calculation 
is carried out in practice. 
Accordingly we use in this appendix the notation $\bm{G}$ 
instead of $\bm{w}$ for the matrix that describes the 
resistor network, and $\sigma$ instead of $D$ for its conductivity. 
We define a vector $\bm{V}=\{V_n\}$, where $V_n$ is the voltage 
at node $n$, analogous to $p_n$. We also define a vector $\bm{I}=\{I_n\}$ 
of injected currents. The Kirchhoff equation \Eq{e1} for a steady 
state can be written as $\bm{G} \bm{V}=0$.

If the nodes were connected to external ``reservoirs" 
the Kirchhoff equation would takes the form $\bm{G} \bm{V}=\bm{I}$. 
The matrix $\bm{G}$ has an eigenvalue zero which is associated  
with a uniform voltage eigenvector. Therefore, it has 
a pseudo-inverse rather than an inverse, and consequently 
the Kirchhoff equation has a solution if and only 
if the net current is ${\sum_n I_n=0}$.

For the purpose of calculating the conductivity
we add a source ${I_1=-1}$ and a drain ${I_2=1}$.
We select the location of the source (site \#1) 
and the drain (site \#2) away from the endpoints. 
From the solution of the Kirchhoff equation we deduce 
\beq
\sigma [d{=}1] \ \ = \ \ \left[(V_2-V_1)/L\right]^{-1} 
\eeq
where $L$ is the distance between the contacts.

With regard to the quasi-one-dimensional model, 
we take the distance between the contacts to 
be ${L'=N/2}$  and look at the voltage drop along 
an inner segment of length ${L=L'-2b}$, 
to avoid the transients at the contact points.

To find the conductivity in the $d{=}2$ case 
we select contacts points that have 
distance ${L\sim (N/2)^{1/2}}$, and use the formula 
\beq
\sigma [d{=}2] \ \ = \ \ \left[(V_2-V_1)/\ln(L/\ell)\right]^{-1} 
\eeq
where $\ell \sim 1$ is the shift of the measurement point
from the contact point. Here the voltage drop 
is divided by $\ln(L/\ell)$ instead of~$L$, 
reflecting the two-dimensional geometry of the flow.

\section{Calculation of the ERH integral}
\label{aERH}

The calculation of the ERH integral for the random site model 
involved the incomplete $\Gamma$ function \cite{gamma}, 
\beq
\Gamma(\ell{+}1,x) = \int_0^x r^{\ell} \eexp{-r} dr = 
\ell! \ \mbox{EXP}_{\ell}(x)  \ \eexp{-x}
\eeq
We first consider the degenerate Mott model. 
We substitute in \Eq{e31},  
the $w(r,\epsilon)$ of \Eq{e4}, 
and the $\rho(r,\epsilon)$ of \Eq{e7} with \Eq{e71}.
Thanks to the $\delta(\epsilon)$ we are left 
just with a $dr$ integration that is split  
into the domains ${0<r<r_c}$ and ${r>r_c}$. 
Namely, 
\beq
D_{\tbox{ERH}} &=& 
\frac{w_0\Omega_d}{2d}\int_0^{r_c} \eexp{-r_c / \xi} \frac{r^{d+1}}{r_0^d} dr 
\nonumber \\
&&+ \frac{w_0\Omega_d}{2d}\int_{r_c}^\infty \eexp{-r / \xi} \frac{r^{d+1}}{r_0^d} dr 
\nonumber \\
&=& \frac{w_0\Omega_d}{2d} \eexp{-r_c / \xi} \frac{r_c^{d+2}}{d+2} \frac{1}{r_0^d}
\nonumber \\
&&+\frac{w_0\Omega_d}{2d} \frac{\xi^{d+2}}{r_0^d} \Gamma\left(d+2,\frac{r_c}{\xi}\right)
\nonumber \\
&=& \frac{w_0\Omega_d \xi^{d+2}}{2d(d+2)r_0^d}\Gamma\left(d+3,\frac{r_c}{\xi}\right)
\eeq
This leads directly to \Eq{e40} with \Eq{e400}.

Turning to the non-degenerated Mott model
we have to deal with a two dimensional integral $drd\epsilon$
that has, as in the previous case, two domains 
${w>w_c}$ and ${0<w<r_c}$. The two domains are 
separated by the line ${\epsilon+(r/\xi)=\epsilon_c}$.
It is therefore natural to change variables:
\beq
x\ \ &=& \ \ \epsilon+(r/\xi) \\
y\ \ &=& \ \ \frac{1}{2}\left(-\epsilon+(r/\xi)\right) 
\eeq
hence 
\beq
D_{\tbox{ERH}} &=& 
\frac{w_0\Omega_d}{2dr_0^d}\int_0^{\epsilon_c}\! \xi dx \int_{-x/2}^{x/2} \! dy \, \eexp{-\epsilon_c} \left(\xi y + \xi \frac{x}{2}\right)^{d+1} 
\nonumber \\
&& + \frac{w_0\Omega_d}{2dr_0^d}\int_{\epsilon_c}^{\infty} \! \xi dx \int_{-x/2}^{x/2} \! dy \, \eexp{-x} \left(\xi y+\xi \frac{x}{2}\right)^{d+1} 
\nonumber \\
&=& \frac{w_0\Omega_d}{2dr_0^d}\xi^{d+2} \eexp{-\epsilon_c} \frac{\epsilon_c^{d+3}}{(d+2)(d+3)}  
\nonumber \\
&& +\frac{w_0\Omega_d}{2dr_0^d(d+2)}\xi^{d+2} \Gamma\left(d+3,\epsilon_c\right)
\nonumber \\
&=& \frac{w_0\Omega_d \xi^{d+2}}{2d(d+2)(d+3)r_0^d}\Gamma\left(d+4,\epsilon_c\right)
\eeq
This leads directly to \Eq{e47} with \Eq{e400}.

\clearpage


 
\clearpage

\onecolumngrid

\twocolumngrid

\clearpage
\end{document}